\documentclass[pra,twocolumn,showpacs]{revtex4}
\usepackage{bbm}

\begin{document}

\title{Comment on ``Secure direct communication with a quantum one-time pad''}

\date{\today}
\author{Holger Hoffmann}
\author{Kim Bostroem}
\author{Timo Felbinger}

\affiliation{Institut f{\"u}r Physik, Universit{\"a}t Potsdam, 14469 Potsdam, Germany}

\begin{abstract}
In the paper [Phys. Rev. A \textbf{69}, 052319 (2004)], a quantum direct communication
protocol is proposed which is claimed to be unconditionally secure even for the case of
a noisy channel. We show that this is not the case by giving an undetectable attack scheme.
\end{abstract}


\maketitle

In \cite{deng:2004}, Deng and Long propose a quantum direct communication protocol which is
briefly described as follows: Bob creates a batch of qubits, each one randomly prepared in
one of the states $|0\rangle,|1\rangle,|+\rangle,|-\rangle$, where
$|\pm\rangle=\frac1{\sqrt 2}(|0\rangle\pm|1\rangle)$, and sends this batch to Alice.
After Alice has received the batch, she chooses a random subset of the batch and performs
measurements in bases randomly chosen from ${\cal B}_z=\{|0\rangle,|1\rangle\}$ and
${\cal B}_x=\{|+\rangle,|-\rangle\}$. Alice announces publicly the positions of the measured
qubits, her choices of basis, and the results of her measurements. Bob compares the measurement
results with his preparation at those positions where the bases coincide. If the bit error
rate is above a certain acceptable threshold, the conversation is aborted. Otherwise Alice
encodes her message on the remaining qubits by applying the operations $\mathbbm 1$ and
$\sigma_y$ to encode 0 and 1. For security purposes, Alice also encodes check bits on randomly chosen qubits. She sends all qubits back to Bob who then decodes Alice's message by measuring each qubit in the corresponding preparation basis. As a final step, Alice and Bob compare the check bits.

The security proof basically consists of two arguments. First, the authors argue that it is
sufficient to prove the security of the first transmission, where the qubits are sent from
Bob to Alice. Second, the authors argue that their protocol inherits the security of the
BB84 scheme, because the control mechanisms are essentially the same.

The second argument is erroneous because the protocol proposed by the authors provides no
privacy amplification step in contrast to BB84. In fact, this step is essential for the
unconditional security of BB84 in the presence of noise (see e.g \cite{gisin:2002, shor:2000}).
That is to say, if there is noise on the channel, Eve can gain a certain amount of
information without being detected, by hiding her presence in the channel noise. 

A successful attack scheme on the proposed protocol can be given as follows. As usual,
Eve is assumed to be limited only by the laws of quantum mechanics. Say, the bit error
rate of the noisy channel is $r$. Eve replaces the noisy channel by an ideal one, and
then measures a fraction $4r$ of the qubits during the first transmission randomly in
the bases ${\cal B}_z$ and ${\cal B}_x$ and resends them to Alice. There is a 50\% chance for Eve to pick the correct basis and even if she picks the wrong basis she has another 50\% chance of not causing a bit error. Therefore, the bit error rate induced by Eve is $r$, which is indistinguishable from
the bit error rate produced by the channel noise, so that the communication will not be
aborted. Alice then encodes her message and sends the qubits back to Bob. Now Eve captures
exactly those qubits on which she has performed her previous measurements and which have
not been discarded due to the control mechanism. She measures these qubits in the same
bases as before, so that she learns what operations Alice has performed on them. Hence,
she is able to eavesdrop a fraction $4r$ of the message bits without being detected. 
Also the additional check bit comparison does not reveal Eve's action. 

The above attack scheme shows that the proposed protocol, at least in its present form,
is not unconditionally secure for the case of a noisy channel. 

If the protocol were enhanced by classical error correction and privacy amplification,
as in the case of BB84, the feature of direct communication would be lost, because after
privacy amplification Alice and Bob would end up with a shared random sequence and not
with a deterministic message. 

This work is supported by the DFG (project WI 337/9-3), and the EU (project IST-2001-38877).

\end{document}